\documentstyle[pre,aps,epsf,multicol]{revtex}
\begin{document}

\title{Multicomponent binary spreading process}
\author{G\'eza \'Odor}
\address{Research Institute for Technical Physics and Materials Science, \\
H-1525 Budapest, P.O.Box 49, Hungary}    
\maketitle

\begin{abstract}
I investigate numerically the phase transitions of two-component 
generalizations of binary spreading processes in one dimension. 
In these models pair annihilation: $AA\to\emptyset$, $BB\to\emptyset$, 
explicit particle diffusion and binary pair production processes compete 
with each other.
Several versions with spatially different productions have been 
explored and shown that for the cases: $2A\to 3A$, $2B\to 3B$ and 
$2A\to 2AB$, $2B\to 2BA$ a phase transition occurs at zero production 
rate ($\sigma=0$), that belongs to the class of N-component, asymmetric 
branching and annihilating random walks, characterized by the order 
parameter exponent $\beta=2$. In the model with particle production: 
$AB\to ABA$, $BA\to BAB$ a phase transition point can be located at 
$\sigma_c=0.3253$ that belongs to the class of the one-component 
binary spreading processes.
\end{abstract}
\begin{multicols}{2}

One-dimensional, non-equilibrium phase transitions have been found to belong 
to a few universality classes, the most robust of them is the directed 
percolation (DP) class \cite{Dick-Mar,Hin2000}. 
According to the hypothesis of \cite{Jan81,Gras82} all continuous phase 
transitions to single absorbing states in homogeneous, single component 
systems with short ranged interactions belong to this class provided 
there is no additional symmetry and quenched randomness present. 
The most well known exception from the robust DP class is the parity 
conserving (PC) class \cite{Gras84}, where a mod 2 conservation of 
particles happens (example in even offspring branching and annihilating 
random walks (BARWe)) and in multi-absorbing state systems where an 
exact $Z_2$ symmetry is satisfied too \cite{MeOdof}.
There are other classes being explored recently where the total number
of particles is conserved \cite{rossi,munoz,pastor,kree,wij,freit,marq}.

In multi-component systems bosonic field theory \cite{Cardy-Tauber} 
simulations \cite{barw2cikk} and density matrix renormalization group 
analysis \cite{Hoy} have revealed the universality class of the 
generalization of the BARWe class. Hard-core particle exclusion effects 
can change both the dynamic \cite{dimercikk,arw2cikk} and static 
\cite{barw2cikk,Park,Lip,dp2cikk,Kor} behavior of one dimensional systems by 
introducing blockades into the particle dynamics. 
Earlier it was shown that infinite number of conservation laws emerge in 
stochastic deposition-evaporation models of Q-mers in one dimension 
\cite{Dhar,Gryn} that split up the phase space into kinetically 
disconnected sectors. That results in initial condition dependent 
autocorrelation functions.

In \cite{arw2cikk} the two-component generalization of the annihilating 
random walk model was introduced taking into account hard-core repulsion
of particles (2-ARW):
\begin{equation}
AA\stackrel{\lambda}{\to}\emptyset \ \ \ 
BB\stackrel{\lambda}\to\emptyset  \ \ \
A\emptyset\stackrel{d}\leftrightarrow\emptyset A \ \ \  
B\emptyset\stackrel{d}\leftrightarrow\emptyset B \ \ \
AB\not\leftrightarrow BA  
\label{arw2mod}
\end{equation}
(where $\lambda$ and $d$ denote the annihilation and diffusion rates)
and showed that the initial conditions influence the particle density 
(order parameter) decay and the dynamical exponents. 
By adding pair creation processes ($A\stackrel{\sigma}\to 2BA$, 
$B\stackrel{\sigma}\to 2AB$) to this model a continuous phase 
transition occurs at $\sigma=0$ creation rate and two universality 
classes appear depending on the arrangement of the offsprings relative 
to the parent \cite{barw2cikk}. 
Namely if the parent separates the offsprings: $A\stackrel{\sigma}{\to}BAB$ 
(2-BARW2s) the steady state density will be higher than in the case 
when they are created on the same site: $A\stackrel{\sigma}{\to}ABB$ 
(2-BARW2a) for a given branching rate $\sigma$ because in the former 
case they are unable to annihilate with each other.
This results in different off-critical order parameter exponents for 
the symmetric and the asymmetric cases ($\beta_s=1/2$ and $\beta_a=2$).
This is in contrast to the widespread beliefs that bosonic field theory
can well describe reaction-diffusion systems in general.
In the field theoretical version \cite{Cardy-Tauber}, where the
$AB\leftrightarrow BA$ exchange is allowed the critical 
behavior is different. Mean-field like and simulation results 
led Kwon et al. \cite{Park} to the assumption that in one-dimensional 
reaction-diffusion systems a series of new universality classes 
should appear if particle exclusion is present. 

In a recent paper \cite{dp2cikk} I showed that if one adds single 
particle creation to the 2-ARW model:
\begin{equation}
A\stackrel{\sigma}{\longrightarrow} AB \ \ \ \ 
B\stackrel{\sigma}{\longrightarrow} BA
\end{equation}
a continuous phase transition occurs again at $\sigma=0$ and the 
critical exponents coincide with that of the 2-BARW2s model, 
although the the parity of particle number is not conserved. 
Therefore this conservation law that was relevant in case of one 
component systems (PC versus DP class) is irrelevant here. 
In \cite{dp2cikk} I made a hypothesis that in coupled branching 
and annihilating random walk systems of N-types of excluding 
particles with continuous transitions at $\sigma=0$ two universality 
classes exist, those of 2-BARW2s and 2-BARW2a, depending on whether the 
reactants can immediately annihilate (i.e. when similar particles 
are not separated by other type(s) of particle(s)) or not.
These classes differ only by the off-critical exponents, 
while the on-critical ones are the same. This is due to the fact that 
the critical point is at zero branching rate ($\sigma=0$) and 
therefore they are the ones determined for the 2-ARW model 
\cite{arw2cikk,barw2cikk}.

In this paper I extend the investigations to coupled binary production
spreading processes, where new universal behavior has recently 
been reported. Studies on the annihilation fission process 
$2A\to\text{\O}$, $2A\to 3A$, $A\emptyset\leftrightarrow\emptyset A$ 
\cite{HT97,Carlon99,Hayepcpd,Odo00,Odo01} found evidence that there is 
a phase transition in this model that does not belong to any previously known 
universality classes.
This model without the single particle diffusion term -- the so called pair 
contact process (PCP), where pairs of particles can annihilate or create
new pairs -- was introduced originally by Jensen \cite{IJen93l} and 
while the static exponents were found to belong to DP class the spreading 
ones show non-universal behavior. By adding explicit single particle 
diffusion \cite{Carlon99} Carlon et al. introduced the so called PCPD 
particle model. The renormalization group analysis of the corresponding 
bosonic field theory was given by Howard and T\"auber \cite{HT97}.
This study predicted a non-DP class transition, but it could not tell 
to which universality class this transition really belongs.
An explanation  based on symmetry arguments are still missing
but numerical simulations suggest \cite{Odo00,HayeDP-ARW} that the behavior 
of this system can be well described (at least for strong diffusion) 
by coupled sub-systems: single particles performing annihilating random walk 
coupled to pairs ($B$) following DP process: $B\to 2B$, $B\to\text{\O}$. 
The model has two non-symmetric absorbing states: one is completely empty, 
in the other a single particle walks randomly. Owing to this fluctuating 
absorbing state this model does not oppose the conditions of the DP hypothesis.

In the low diffusion region ($d<\sim0.4$) some exponents of the PCPD model 
are close to those of the PC class but the order parameter exponent 
($\beta$) has been found to be very far away from both of the DP and PC 
class values \cite{Odo00}. In fact this system does not exhibit neither 
a $Z_2$ symmetry nor a parity conservation that appear in models
with PC class transition. In the high diffusion region the critical 
exponents seem to be different \cite{Carlon99,Odo00,Grasp} suggesting an 
other universality class there \cite{Odo00}. This is also supported by
the pair mean-field results \cite{Carlon99}. A recent universal finite 
size scaling amplitude study \cite{UFSSA} suggests however that a single 
universality class with strong corrections to scaling may also be possible.

It is conjectured  by Henkel and Hinrichsen \cite{HH} that this kind of 
phase transition appears in models where (i) solitary particles diffuse, 
(ii) particle creation requires two particles and (iii) particle removal 
requires at least two particles to meet.
Very recently Park et al. \cite{binary} have investigated the
parity conserving version of the PCPD model 
($2A\to 4A$, $2A\to\emptyset$, $A\emptyset\leftrightarrow\emptyset A$) 
and contrary the apparent conservation law they have found similar 
scaling behavior that led them to the assumption that the binary 
nature of the offspring production is a necessary condition for 
this class. Other conditions that would influence the occurrence of this class
should be clarified too. In this paper I address the question whether the
particle exclusion effects are relevant like in the case of BARW processes
and whether the hypothesis set up for N-BARW systems \cite{dp2cikk} could
be extended.

One site update step of the applied algorithms consist of the following 
processes. A particle is selected randomly. A left or right nearest 
neighbor is chosen with probability 0.5. With probability $\sigma$ a 
pair production is attempted in case of an appropriate neighbor. 
Otherwise (with probability $d=\lambda=1-\sigma$) a hopping is attempted 
if the neighbor is empty or if the neighbor if filled with a particle of same 
type they are annihilated. The following models with the same diffusion and 
annihilation terms as (\ref{arw2mod}) and different production processes 
will be investigated here.

a) Production and annihilation random walk model (2-PARW):
\begin{eqnarray}
AA\stackrel{\sigma/2}{\longrightarrow}AAB, \ \ \ 
AA\stackrel{\sigma/2}{\longrightarrow}BAA, \\
BB\stackrel{\sigma/2}{\longrightarrow}BBA, \ \ \
BB\stackrel{\sigma/2}{\longrightarrow}ABB \ .
\end{eqnarray}

b) Symmetric production and annihilation random walk model 
(2-PARWS):
\begin{eqnarray}
AA\stackrel{\sigma}{\longrightarrow}AAA, \\ 
BB\stackrel{\sigma}{\longrightarrow}BBB \ .
\end{eqnarray}

c) Asymmetric  production and annihilation random walk model
(2-PARWA):
\begin{eqnarray}
AB\stackrel{\sigma/2}{\longrightarrow}ABB, \ \ \
AB\stackrel{\sigma/2}{\longrightarrow}AAB, \\ 
BA\stackrel{\sigma/2}{\longrightarrow}BAA, \ \ \
BA\stackrel{\sigma/2}{\longrightarrow}BBA \ .
\end{eqnarray}

d) Asymmetric production and annihilation random walk model
with spatially symmetric creation (2-PARWAS):
\begin{eqnarray}
AB\stackrel{\sigma/2}{\longrightarrow}ABA, \ \ \
AB\stackrel{\sigma/2}{\longrightarrow}BAB, \\ 
BA\stackrel{\sigma/2}{\longrightarrow}BAB, \ \ \
BA\stackrel{\sigma/2}{\longrightarrow}ABA \ .
\end{eqnarray}

The evolution of particle densities were followed by Monte-Carlo 
simulations started from randomly distributed $A,B,\emptyset$ sites
in systems of sizes $L=10^5$ and periodic boundary conditions.

The 2-PARWA (c) model does not have an active steady state. The $AA$ and $BB$
pairs annihilate themselves on contact, while if an $A$ and $B$ particle 
meets an $AB\to ABB\to A$ process reduces out blockades therefore the 
densities decay with the $\rho\propto t^{-1/2}$ law for $\sigma > 0$. 
This was confirmed by my simulations. Note that for $\sigma=0$ the blockades
persist and in case of random initial state a $\rho\propto t^{-1/4}$ decay 
can be observed \cite{barw2cikk}.

\begin{figure}
\epsfxsize=70mm
\epsffile{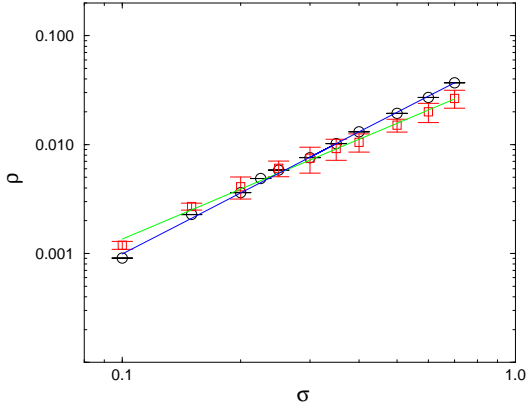}
\vspace{4mm}
\caption{Steady state densities as the function of $\sigma$ in the 2-PARW 
(squares) and in the 2-PARWS (circles) models.}
\label{parwll}
\end{figure}

The 2-PARW (a) and the 2-PARWS (b) models exhibit active steady states 
for $\sigma>0$ with a continuous phase transition at $\sigma=0$. Therefore
the exponents at the critical point will be those of the ARW-2 model.  
The convergence to the steady state is very slow. For $\sigma=0.1$ it was
longer than $10^9$ Monte Carlo steps (MCS). 
This has limited the simulations by approaching the critical point at 
$\sigma=0$. However as Figure \ref{parwll} shows a rather good scaling 
behavior of the density versus $\sigma$ can be observed.
The local slope analysis shows that the scaling behavior extrapolates to
$\beta=2.1(2)$ in case of the 2-PARWS model and to $\beta=1.9(2)$ in 
case of the 2-PARW model. These values are in agreement with those of
the 2-BARW2a class ($\beta=2$), where production is such that pair 
annihilation is enhanced.
\begin{figure}
\epsfxsize=80mm
\epsfysize=90mm
\epsffile{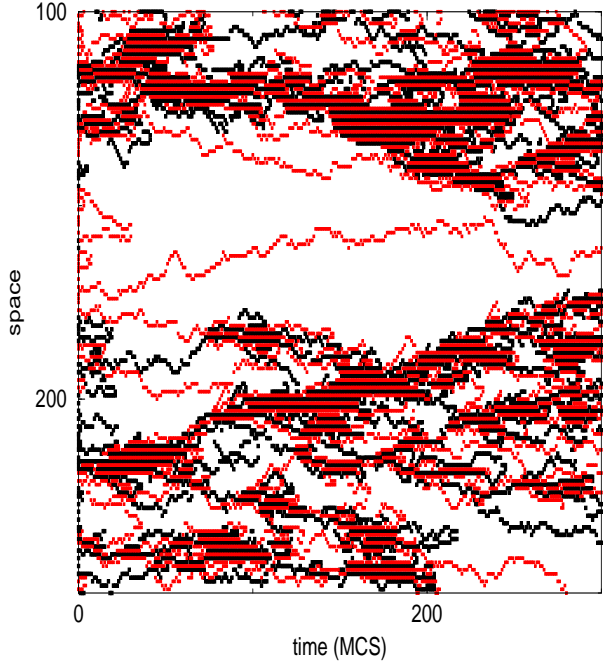}
\caption{Space-time evolution from random initial state of the 2-PARWAS 
model at the critical point. Black dots correspond to $A$ particles, 
red dots to $B$-s.}
\label{rajz}
\end{figure}
In case of the 2-PARWAS (d) model the $AB$ blockades proliferate by 
production events. As the consequence of this an active steady state 
appears for $\sigma > 0.3253(1)$ with a continuous phase transition.
The space-time evolution from random initial state shows (Fig.\ref{rajz})
that compact domains of alternating $..ABAB..$ sequences separated by
lonely wandering particles are formed.
This is very similar to what was seen in case of one-component binary 
spreading processes \cite{HayeDP-ARW}: compact domains within a cloud of
lonely random walkers, except that now domains are built up from 
alternating sequences only. This means that $..AAAA...$ and $...BBBB...$ 
domains decay at this annihilation rate and particle blocking is responsible 
for the compact clusters. In the language of coupled DP + ARW model 
\cite{HayeDP-ARW} the pairs following DP process are the $AB$ pairs 
now, which cannot decay spontaneously but through an annihilation process: 
$AB + BA\to\emptyset$. They interact with two types of particles 
executing annihilating random walk with exclusion.
 
Simulations from random initial state were run up to $10^6$ MCS. 
The local slopes of the particle density decay
\begin{equation}
\alpha_{eff}(t) = {- \ln \left[ \rho(t) / \rho(t/m) \right] 
\over \ln(m)} \label{slopes}
\end{equation}
(where $m=8$ is used) at the critical point go to 
exponent $\alpha$ by a straight line asymptotically, while in 
sub(super)-critical cases they veer down(up) respectively. 
\begin{figure}
\epsfxsize=70mm
\epsffile{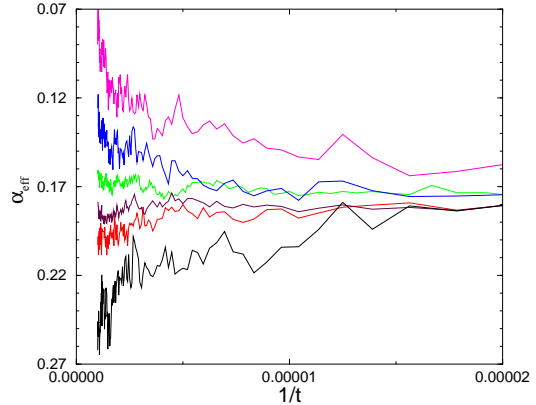}
\vspace{2mm}
\caption{Local slopes of the density decay in the PARWAS model.
Different curves correspond to $\sigma=0.325$, $0.3252$, $0.3253$ 
$0.3254$, $0.3255$, $0.326$ (from bottom to top).}
\label{dec}
\end{figure}
At the critical point ($\sigma_c=0.3253(1))$ one can estimate that
the effective exponent tends to $\alpha = 0.19(1)$, which is higher
than the the exponent of the 1+1 dimensional directed percolation 
0.1595(1) \cite{IJensen} and in fairly good agreement with that of the
PCPD model in the high diffusion rate region (0.20(1)) \cite{Odo00}.

\begin{figure}
\epsfxsize=70mm
\epsffile{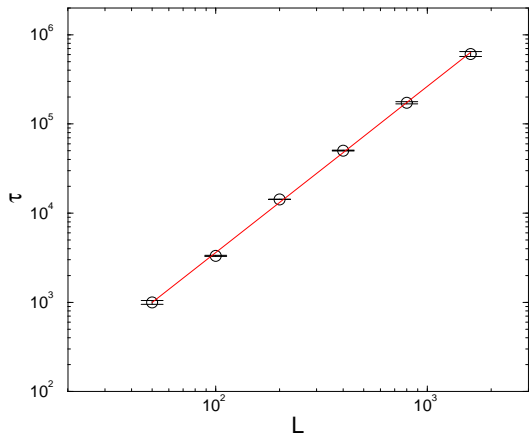}
\vspace{4mm}
\caption{Lifetime versus system size at the critical point}
\label{tau}
\end{figure}

In the supercritical region the steady states have been determined for 
different $\epsilon=\sigma-\sigma_c$ values. Following level-off the 
densities were averaged over $10^4$ MCS and $1000$ samples.
By looking at the effective exponent defined as
\begin{equation}
\beta_{eff}(\epsilon_i) = \frac {\ln \rho(\epsilon_i) -
\ln \rho(\epsilon_{i-1})} {\ln \epsilon_i - \ln \epsilon_{i-1}} \ \ ,
\end{equation}
one can read-off: $\beta_{eff}\to\beta \simeq 0.37(2)$, which is again
higher than that of the 1+1 dimensional DP value 0.27649(4) \cite{Dick-Jen}, 
and agrees with that of the PCPD model in the high diffusion rate region 
(0.39(2)) \cite{Odo00}.
\begin{figure}
\epsfxsize=70mm
\epsffile{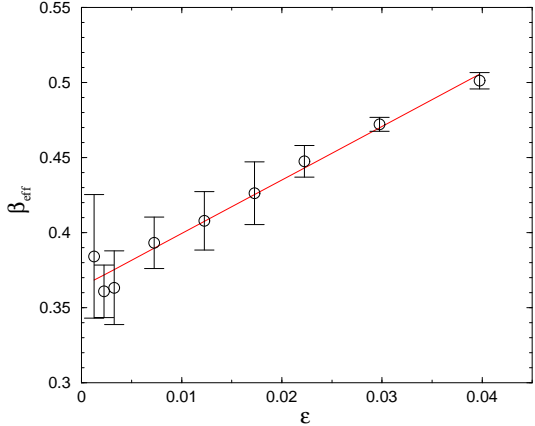}
\vspace{4mm}
\caption{Effective order parameter exponent results. Linear extrapolation 
results in $\beta=0.37(2)$.}
\label{beta}
\end{figure}
Finally the survival probability ($P(t)$) of systems started from 
random initial condition was measured for sizes: $L=50,100,200,400,800,1600$. 
The characteristic time $\tau(L)$ to decay to $P(\tau)=0.9$ 
was determined and shown on Figure \ref{tau}. 
At criticality one expects the finite size scaling
\begin{equation}
\tau(L) \propto L^{Z} \ ,
\end{equation}
where $Z$ is the dynamical exponent. The power-law fitting resulted in
$Z=1.81(2)$, which is far away from the DP value $Z=1.580740(34)$ 
\cite{IJensen} but close to various estimates for the PCPD value
$Z=1.75(10)$ \cite{Carlon99,Hayepcpd}.

In conclusion I have shown in this work that the hypothesis that I made for
N-BARW models with exclusion \cite{dp2cikk} may be extended for coupled 
binary production annihilation models. The critical point in the 2-PARW 
and 2-PARWS models, where $AA$ and $BB$ pairs create offsprings, continuous
phase transition occurs at $\sigma=0$ production rate therefore 
the on-critical exponents coincide with those of the the 2-ARW model. 
The simulations for the off-critical behavior of the order parameter 
have shown that the transition belongs to the 2-BARW2a class. 
The robustness of this class is striking especially in case of the 
2-PARWS model where in principle two copies of PCPD models are 
superimposed and coupled with the exclusion interaction only.

If the production is generated by different types of particles ($AB$)
such that alternating sequences are generated (2-PARWAS model) 
the space-time evolution will resemble to the of the PCPD model 
with alternating frozen sequences inside the compact domains. 
This system exhibits a continuous phase transition at 
$\sigma=0.3253(1)$ with exponents in fairly good agreement with 
those of the PCPD model in the high diffusion region. 
In the model where $AB$ pairs create offsprings in a such a way that 
prompt annihilation is possible active steady states are not formed for 
any $\sigma$ and the density decays without blockades
for $\sigma>0$ as $\rho\propto t^{-0.5}$ but a crossover to 
2-ARW model scaling $\rho\propto t^{-0.25}$ occurs at $\sigma=0$.

\noindent
{\bf Acknowledgements:}\\
The author thanks H. Chat\'e and P. Grassberger for their comments.
Support from Hungarian research funds OTKA (No. T-25286) and Bolyai 
(No. BO/00142/99) is acknowledged. The simulations were performed on 
the parallel cluster of SZTAKI and on the supercomputer of NIIF Hungary.

\end{multicols}

\begin{references}

\bibitem{Dick-Mar} For references see : J.~Marro and R.~Dickman,
\newblock {\em Nonequilibrium phase transitions in lattice models},
\newblock Cambridge University Press, Cambridge, 1999.
\bibitem{Hin2000} H.~Hinrichsen, Adv. Phys. {\bf 49}, 815 (2000).
\bibitem{Jan81} H. K. Janssen, Z. Phys. B {\bf 42}, 151 (1981).
\bibitem{Gras82} P. Grassberger, Z. Phys. B {\bf 47}, 365 (1982).
\bibitem{Gras84} P.~Grassberger, F.~Krause and T.~ von der
Twer, J. Phys. A:Math.Gen., {\bf 17}, L105 (1984).
\bibitem{MeOdof} For an overview and references see: 
N. Menyh\'ard and G. \'Odor, Brazilian J. of Physics {\bf 30}, 113 (2000). 
\bibitem{rossi} M. Rossi, R. Pastor-Satorras and A. Vespignani, Phys. Rev. 
Lett. {\bf 85}, 1803 (2000).
\bibitem{munoz} M. A. Munoz, R. Dickman, A. Vespignani and S. Zapperi,
Phys. Rev. {\bf 59}, 6175 (1999).
\bibitem{pastor} R. Pastor-Satorras and A. Vespignani, Phys. Rev. E 
{\bf 62}, 5875 (2000).
\bibitem{kree} R. Kree, B. Schaub and B. Schmitmann, Phys. Rev. A {\bf 39},
2214 (1989).
\bibitem{wij} F. van Wijland, K. Oerding and H. J. Hilhorst, Physica A
{\bf 251}, 179 (1998).
\bibitem{freit} J. de Freitas, L. S. Lucena, L. R. da Silva and 
H. J. Hilhorst, Phys. Rev. E {\bf 61}, 6330 (2000).
\bibitem{marq} M. C. Marques, Phys. Rev. E {\bf 64}, 016104-1 (2001).
\bibitem{Cardy-Tauber} J. L. Cardy and U. C. T\"auber, J. Stat. Phys.
{\bf 90}, 1 (1998).
\bibitem{barw2cikk} G. \'Odor, Phys. Rev. E {\bf 63}, 021113 (2001).
\bibitem{Hoy} J. Hooyberghs, E. Carlon and C. Vancderzande, 
Phys. Rev. E {\bf 64}, 036124 (2001).
\bibitem{dimercikk} H. Hinrichsen and G. \'Odor, Phys. Rev. E {\bf 60}, 
3842 (1999).
\bibitem{arw2cikk} G. \'Odor and N. Menyh\'ard, Phys. Rev. E. 
{\bf 61}, 6404 (2000)
\bibitem{Park} S. Kwon, J. Lee and H. Park, Phys. Rev. Lett. 
{\bf 85}, 1682 (2000).
\bibitem{Lip} A. Lipowski and M. Droz, Phys. Rev. E {\bf 64} (2001) 031107.
\bibitem{dp2cikk} G. \'Odor, Phys. Rev. E {\bf 63}, 0256108 (2001).
\bibitem{Kor} S. Kwon and H. Park, cond-mat/0010380.
\bibitem{Dhar} D. Dhar and M. Barma, Pramana {\bf 41}, L193 (1993).
\bibitem{Gryn} R. B. Stinchcomb, M. D. Grynberg and M. Barma, 
Phys. Rev. E {\bf 47}, 4018 (1993).
\bibitem{HT97} M.~J. Howard and U.~C. T{\"a}uber,  
{J. Phys.} {\bf A 30}, 7721 (1997).
\bibitem{Carlon99} E.~Carlon, M.~Henkel and U.~Schollw{\"o}ck,
Phys. Rev. E {\bf 63}, 036101-1 (2001).
\bibitem{Hayepcpd} H.~Hinrichsen, Phys. Rev. E {\bf 63}, 036102-1 (2001).
\bibitem{Odo00} G. \'Odor , Phys. Rev. E {\bf 62}, R3027 (2000).
\bibitem{Odo01} G. \'Odor , Phys. Rev. E {\bf 63}, 067104 (2001).
\bibitem{Grasp} P. Grassberger, private communication.
\bibitem{UFSSA} M. Henkel and U. Schollw\"ock, J. Phys. A {\bf 34}, 3333 (2001).
\bibitem{IJen93l} I. Jensen, Phys. Rev. Lett. {\bf 70}, 1465 (1993).
\bibitem{HayeDP-ARW} H.~Hinrichsen, Physica A {\bf 291}, 275-286 (2001).
\bibitem{HH} M.~Henkel and H.~Hinrichsen, J. Phys. A {\bf 34}, 1561 (2001).
\bibitem{binary} K. Park, H. Hinrichsen, and In-mook Kim, Phys. Rev. E 
{\bf 63}, 065103(R) (2001).
\bibitem{IJensen} I. Jensen, Phys. Rev. Lett. {\bf 77}, 4988 (1996).
\bibitem{Dick-Jen} R. Dickman and I. Jensen, 
Phys. Rev. Lett. {\bf 67}, 2391 (1991).

\end{references}
\end{document}